\documentclass[aps,prl,twocolumn,showpacs,byrevtex]{revtex4}
\usepackage{times,mathptm}% for making `pdf' file
\usepackage[dvips]{graphicx,color}% for inclusion of graphics file(s)
\DeclareGraphicsExtensions{.pdf,.png,.jpg,.eps}

\begin{document}
\title{Nature of the Insulating Ground State of the 5$d$ Post-Perovskite CaIrO$_3$}
\author{Sun-Woo Kim$^{1}$, Chen Liu$^{2}$, Hyun-Jung Kim$^{1,3}$, Jun-Ho Lee$^{1,4}$, Yongxin Yao$^{2,{\dagger}}$, Kai-Ming Ho$^{2,3}$, and Jun-Hyung Cho$^{1,3,*}$}
\affiliation{$^1$ Department of Physics and Research Institute for Natural Sciences, Hanyang University,
17 Haengdang-Dong, Seongdong-Ku, Seoul 133-791, Korea \\
$^2$ Ames Laboratory and Department of Physics and Astronomy, Iowa State University, Ames, Iowa 50011, USA\\
$^3$ International Center for Quantum Design of Functional Materials (ICQD), HFNL,
University of Science and Technology of China, Hefei, Anhui 230026, China\\
$^4$ Korea Institute for Advanced Study, 85 Hoegiro, Dongdaemun-gu, Seoul 130-722, Korea}
\date{\today}

\begin{abstract}

The insulating ground state of 5$d$ transition metal oxide CaIrO$_3$ has been classified as a Mott-type insulator. Based on a systematic density functional theory (DFT) study with local, semilocal, and hybrid exchange-correlation functionals, we reveal that the Ir $t_{2g}$ states exhibit large splittings and one-dimensional electronic states along the $c$ axis due to a tetragonal crystal field. Our hybrid DFT calculation adequately describes the antiferromagnetic (AFM) order along the $c$ direction via a superexchange interaction between Ir$^{4+}$ spins. Furthermore, the spin-orbit coupling (SOC) hybridizes the $t_{2g}$ states to open an insulating gap. These results indicate that CaIrO$_3$ can be represented as a spin-orbit Slater insulator, driven by the interplay between a long-range AFM order and the SOC. Such a Slater mechanism for the gap formation is also demonstrated by the DFT + dynamical mean field theory calculation, where the metal-insulator transition and the paramagnetic to AFM phase transition are concomitant with each other.
\end{abstract}

\maketitle
%\begin{multicols}{2}

%\vspace{0.4cm}
%\section{I. INTRODUCTION}
%\vspace{0.4cm}

One of the most important phenomena in condensed matter physics is the Mott transition driven by electron-electron correlations~\cite{Mott,Imada}. In 3$d$ transition-metal oxides (TMOs), the localized 3$d$ orbitals are responsible for the strong on-site Coulomb repulsion ($U$), leading to a Mott-Hubbard insulator where $U$ splits a half-filled band into lower and upper Hubbard bands. Surprisingly, despite weaker $U$ in 5$d$ TMOs due to the very delocalized 5$d$ orbitals, a series of Ir oxides such as Sr$_2$IrO$_4$~\cite{BJKim,BJKim2,SJMoon,Arita,li}, Na$_2$IrO$_3$~\cite{Shitade,Comin,Gretarsson}, and CaIrO$_3$~\cite{Ohgushi2,Subedi,Bogdanov,Ohgushi,Sala} including Ir$^{4+}$ ions with five valence electrons exhibit an insulating ground state. For this unusual insulating behavior of the 5$d$ iridates, it was proposed that spin-orbit coupling (SOC) splits the Ir $t_{2g}$ states into completely filled $j_{\rm eff}$ = 3/2 bands and a narrow half-filled $j_{\rm eff}$ = 1/2 band at the Fermi level ($E_F$), and the latter band is further split into two Hubbard subbands by moderate Coulomb repulsion~\cite{BJKim,BJKim2,SJMoon}. Such a $j_{\rm eff}$ = 1/2 Mott-Hubbard scenario has, however, been challenged by an alternative scenario of Slater mechanism~\cite{Slater} based on the single-particle band picture, where the opening of insulating gap in 5$d$ TMOs is driven by a long-range magnetic ordering~\cite{Arita,li,Calder,HJKim}.

Here we focus on the post-perovskite CaIrO$_3$ with a highly anisotropic geometry where IrO$_6$ octahedra share corners along the $c$ axis and have common edges along the $a$ axis (see Fig. 1). Recently, the nature of the ground state in CaIrO$_3$ has been an object of hot debate~\cite{Ohgushi2,Ohgushi,Sala,Subedi,Bogdanov}. On the basis of resonant x-ray magnetic scattering (RMXS) experiment, Ohgushi $et$ $al.$~\cite{Ohgushi} claimed the robustness of the $j_{\rm eff}$ = 1/2 ground state against structural distortions. However, a resonant inelastic x-ray scattering (RIXS) experiment of Sala $et$ $al$.~\cite{Sala} concluded that CaIrO$_3$ is not a $j_{\rm eff}$ = 1/2 iridate by showing that the $j_{\rm eff}$ = 1/2 state is severely altered by a large tetragonal crystal field splitting, therefore proposing the existence of Mott insulator beyond the $j_{\rm eff}$ = 1/2 ground state. On the theoretical side, using density-functional theory (DFT) calculations within the local density approximation (LDA) including SOC, Subedi~\cite{Subedi} interpreted the $t_{2g}$ states in terms of the $j_{\rm eff}$ = 1/2 and $j_{\rm eff}$ = 3/2 states, and further showed that the introduction of $U$ splits the $j_{\rm eff}$ = 1/2 bands into fully occupied lower and unoccupied upper Hubbard bands, thereby supporting the Mott-Hubbard scenario. Contrasting with the $j_{\rm eff}$ = 1/2 ground state generated by equally hybridizing the three $d_{xy}$, $d_{yz}$, and $d_{zx}$ orbitals~\cite{Ohgushi}, $ab$ $intio$ wave-function quantum chemical calculation~\cite{Bogdanov} predicted the highly uneven admixture of $xy$, $yz$, and $zx$ characters for the relativistic $t_{2g}$ states. Therefore, previous experimental and theoretical studies of CaIrO$_3$ have not reached a consensus on the presence of the $j_{\rm eff}$ = 1/2 ground state, but concluded in the same way that CaIrO$_3$ belongs to a Mott-Hubbard insulator. Despite such a contradiction for the nature of the ground state, it is well established~\cite{Ohgushi,Sala,Subedi,Bogdanov} that the insulating ground state of CaIrO$_3$ exhibits the stripe-type magnetic order with a strong antiferromagnetic (AFM) coupling along the $c$ axis and a weak ferromagnetic one along the $a$ axis (hereafter designated as the AFM structure).
\begin{figure}[ht]
\centering{ \includegraphics[width=7.7cm]{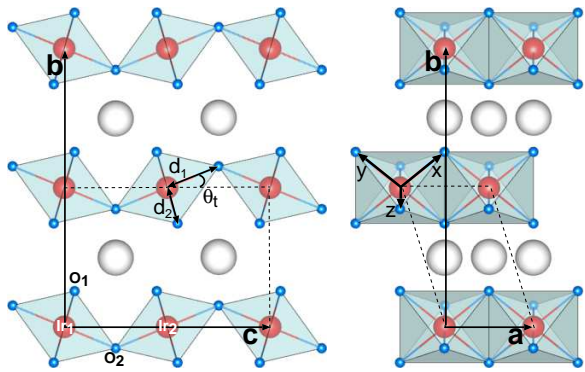} }
\caption{(Color online) Crystal structure of CaIrO$_3$ projected on the $bc$ plane (left side) and the $ab$ plane (right side). $a$, $b$, and $c$ denote unit vectors of the conventional unit cell. The primitive unit cell is indicated by the dashed lines. The large, medium, and small circles represent Ca, Ir, and O atoms, respectively. The bond lengths ($d_1$ and $d_2$) and the tilt angle (${\theta}_t$) are indicated for Table IIS. The reference frame $xyz$ is drawn in an IrO$_6$ octahedron.}
\end{figure}

In this Letter, we investigate the nature of the ground state of CaIrO$_3$ by using comprehensive DFT calculations with local, semilocal, and hybrid exchange-correlation functionals as well as by including dynamical mean field theory (DMFT). We find that the $t_{2g}$ states are significantly split by a compression of IrO$_6$ octahedra along the $c$ axis and particularly two $t_{2g}$ states (designated as $t_{2g}^{S1}$ and $t_{2g}^{S2}$) have dominant $d_{yz}$ and $d_{zx}$ characters with large band dispersions, indicating one-dimensional (1D) electronic states along the $c$ axis. Our hybrid DFT calculation adequately describes the delocalized $t_{2g}^{S1}$ and $t_{2g}^{S2}$ states to stabilize the AFM order along the $c$ axis via superexchange interaction between Ir$^{4+}$ spins. Moreover, the SOC is found to hybridize the $t_{2g}^{S1}$ and $t_{2g}^{S2}$ states with other $t_{2g}$ states to open an insulating gap. It is thus elucidated that the gap formation in CaIrO$_3$ is driven by the interplay between a long-range AFM order and the SOC, representing a spin-orbit Slater insulator. This Slater mechanism for the gap formation in CaIrO$_3$ is further supported by the DFT+DMFT calculation.

We first study the ground state of CaIrO$_3$ using the LDA, GGA, and hybrid DFT calculations~\cite{method} without SOC. Interestingly, the different ground states are predicted depending on the employed exchange-correlation functionals: i.e., the CA functional predicts the nonmagnetic (NM) ground state, while the PBE and HSE functionals favor the AFM structure over the NM structure by ${\Delta}E_{\rm NM-AFM}$ = 21.1 and 85.8 meV per primitive unit cell, respectively. The calculated structural parameters of Ir-O bond lengths ($d_{1}$ and $d_{2}$ in Fig. 1) and tilt angle (${\theta}_t$ in Fig. 1) change little depending on the employed exchange-correlation functionals. Our values of $d_{1}$/$d_{2}$ ${\approx}$ 0.96 and ${\theta}_t$ ${\approx}$ 23$^\circ$ indicate that the IrO$_6$ octahedra are compressed along the $c$ axis with a large tilt, in good agreement with an XRD analysis~\cite{Hirai} and a previous LDA calculation~\cite{Subedi} (see Table IIS of the Supplemental Material~\cite{Supple}). Figure 2(a) shows the LDA band structure of the NM structure together with $d$-orbitals projected bands and the charge characters of the $t_{2g}$ states. There are six $t_{2g}$ bands originating from two different Ir atoms within the primitive unit cell, which are grouped into two doublets ($t_{2g}^{D1}$ and $t_{2g}^{D2}$) and two singlets ($t_{2g}^{S1}$ and $t_{2g}^{S2}$). We find the presence of partially occupied $t_{2g}$ states crossing $E_{F}$, indicating a metallic feature. It is seen in Fig. 2(a) that (i) the $t_{2g}^{D1}$ and $t_{2g}^{D2}$ doublets are significantly separated by ${\sim}$1.5 eV at the ${\Gamma}$ point and (ii) the low(high)-lying $t_{2g}^{D2}$ ($t_{2g}^{D1}$) states involve dominantly $d_{xy}$ ($d_{yz}$ and $d_{zx}$) component(s). These different orbital characters of $t_{2g}^{D1}$ and $t_{2g}^{D2}$ can be attributed to a tetragonal crystal field due to the compressed IrO$_6$ octahedra along the $c$ axis. In addition, the $t_{2g}^{S1}$ and $t_{2g}^{S2}$ states also have dominant $d_{yz}$ and $d_{zx}$ characters with a large band width of ${\sim}$2.5 eV, indicating 1D electronic states along the $c$ axis. As shown in the inset in Fig. 2(a), the charge character of $t_{2g}^{S1}$ (or $t_{2g}^{S2}$) shows a strong hybridization with O 2$p$ orbitals along the $c$ axis, reflecting its 1D band feature. This 1D features of the $t_{2g}^{S1}$ and $t_{2g}^{S2}$ states play an important role in determining the electronic and magnetic properties of CaIrO$_3$, as discussed below.

\begin{figure*}[t]
\centering{ \includegraphics[width=15cm]{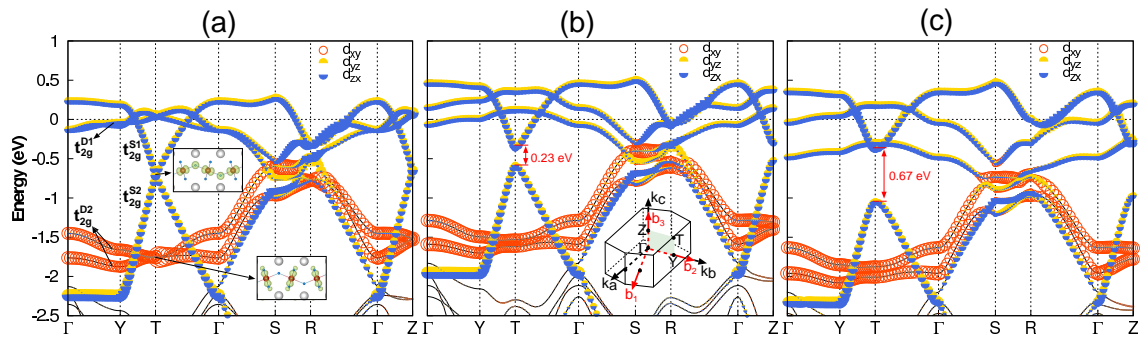} }
\caption{(Color online) Band structures of (a) the NM structure obtained using LDA, (b) the AFM structure obtained using PBE, and (c) the AFM structure obtained using HSE. The band dispersions are plotted along the path ${\Gamma}$(0,0,0) ${\rightarrow}$ $Y$(0,0.5,0) ${\rightarrow}$ $T$(0,0.5,0.5) ${\rightarrow}$ ${\Gamma}$(0,0,0) ${\rightarrow}$ $S$(0.25,0.25,0) ${\rightarrow}$
$R$(0.25,0.25,0.5) ${\rightarrow}$ ${\Gamma}$(0,0,0) ${\rightarrow}$ $Z$(0,0,0.5) in the Brillouin zone of the unit cell [see the inset in (b)]. The energy zero represents the Fermi level. The bands projected onto $d_{xy}$, $d_{yz}$, and $d_{zx}$ orbitals are also displayed. Here, the radii of circle and semicircles are proportional to the weights of corresponding orbitals. In (a), the charge characters of $t_{2g}^{S1}$ and $t_{2g}^{D2}$ states at the $T$ point are shown with an isosurface of 0.02 $e$/{\AA}$^3$.}
\end{figure*}

Figure 2(b) and 2(c) show the PBE and HSE band structures of the AFM structure, respectively. Here, the overall band dispersion of the AFM structure is similar to that [Fig. 2(a)] of the NM structure. However, there are two conspicuous changes in the AFM band structure: i.e., one is a gap opening $E_{g}^{S1-S2}$ of the $t_{2g}^{S1}$ and $t_{2g}^{S2}$ singlets at the $T$ point and the other is some separation of the two bands in the $t_{2g}^{D1}$ doublet. We obtain $E_{g}^{S1-S2}$ = 0.23 eV [Fig. 2(b)] and 0.67 eV [Fig. 2(c)] from PBE and HSE, respectively. To understand the microscopic mechanism for these changes in the AFM band structure, we plot the spin-polarized $d$-orbitals projected bands (see Fig. 5S of the Supplemental Material~\cite{Supple}). We find that the $t_{2g}^{S1}$ and $t_{2g}^{S2}$ states (or the two states in $t_{2g}^{D1}$) with the same spin direction, localized at two different Ir sites, hybridize with each other, yielding an energy gain from the exchange kinetic energy and the formation of $E_{g}^{S1-S2}$~\cite{Sato}. This so-called superexchange interaction~\cite{Goodenough,Kanamori} between Ir$^{4+}$ spins results in the stabilization of the AFM order along the $c$ axis. Our finding that the $t_{2g}^{S1}$, $t_{2g}^{S2}$, and $t_{2g}^{D1}$ states with $d_{yz}$ and $d_{zx}$ characters are associated with the AFM order does not support the interpretation of previous RMXS experiment~\cite{Ohgushi} that the $j_{\rm eff}$ = 1/2 state stabilizes the AFM order. In Table I, we list the calculated magnetic moments of Ir and O atoms in the AFM structure. Our HSE calculation gives a spin magnetic moment $m_S$ = 0.48 (0.07) ${\mu}_B$ for Ir (O$_1$) atoms, larger than $m_S$ = 0.31 (0.05) ${\mu}_B$ computed using PBE. We note that the LDA and GGA tend to stabilize artificially delocalized electronic states due to their inherent self-interaction error because delocalization reduces the spurious self-repulsion of electrons~\cite{SIE1,SIE2}. This aspect of LDA and GGA may account for why our LDA calculation predicts a metallic NM ground state and our PBE values of ${\Delta}E_{\rm NM-AFM}$ and $m_S$ are relatively smaller than the corresponding HSE ones (see Table I).

Next, we examine the effect of SOC on the geometry and band structure of the AFM structure using PBE+SOC and HSE+SOC~\cite{method}. As shown in Table IIS of the Supplemental Material, the inclusion of SOC changes little $d_{1}$/$d_{2}$ and ${\theta}_t$ by less that 0.01 and 1$^\circ$, respectively. Figures 3(a) and 3(b) show the PBE+SOC and HSE+SOC band structures of the AFM structure, respectively. Compared to the PBE [Fig. 2(b)] and HSE [Fig. 2(c)] results, $d$-orbitals projected bands clearly show that the inclusion of SOC does not change the $d_{xy}$, $d_{yz}$, and $d_{zx}$ characters of the $t_{2g}^{D1}$, $t_{2g}^{D2}$, $t_{2g}^{S1}$ and $t_{2g}^{S2}$ states, but gives rise to their strong hybridizations leading to well-separated $t_{2g}$ bands with hybridization gaps. This result doesn't support the $j_{\rm eff}$ = 1/2 ground-state picture proposed by RMXS experiment~\cite{Ohgushi}, where SOC can dramatically affect the $t_{2g}$ states to produce the $j_{\rm eff}$ = 1/2 state by equally mixing up the $d_{xy}$, $d_{yz}$, and $d_{zx}$ orbitals. We note that for the $t_{2g}$ state near $E_F$, the partial densities of states (PDOS) of the $d_{xy}$, $d_{yz}$, and $d_{zx}$ orbitals obtained using HSE give a ratio of 0\%:50\%:50\%, but change into 8\%:46\%:46\% for the HSE+SOC calculation [see Fig. 6S of the Supplemental Material]. The latter PDOS ratio of $d_{xy}$, $d_{yz}$, and $d_{zx}$ is very consistent with the interpretation of RIXS experiment~\cite{Sala} that the $t_{2g}$ state near $E_F$ was estimated as ${\mp}0.32|xy,{\mp}>$ + 0.67 (($|yz,{\pm}>{\mp}i|zx,{\pm}>$). As a consequence of the SOC-induced hybridization, PBE+SOC still does not open an insulating gap [Fig. 3(a)], but HSE+SOC opens $E_g$ = 0.32 eV [Fig. 3(b)], close to the experimental data of ${\sim}$0.34 eV~\cite{Ohgushi2}. Especially, the DOS [see Fig. 3(b)] obtained using HSE+SOC explains well the two $d$-$d$ interband transitions observed in RIXS experiment~\cite{DOS}. It is noteworthy that the gap opening in HSE+SOC can be realized by the strong interplay between a long-range AFM order and the SOC: i.e., the enhanced separation of the $t_{2g}^{S1}$ and $t_{2g}^{D1}$ bands in the AFM structure [see Fig. 2(c)] and the SOC-induced hybridization are combined to yield the opening of an insulating gap. Thus, we can say that the single-particle band theory within the HSE+SOC scheme predicts well the magnetic-insulating ground state of CaIrO$_3$, being represented as a spin-orbit Slater insulator.

\begin{figure*}[t]
\centering{ \includegraphics[width=15cm]{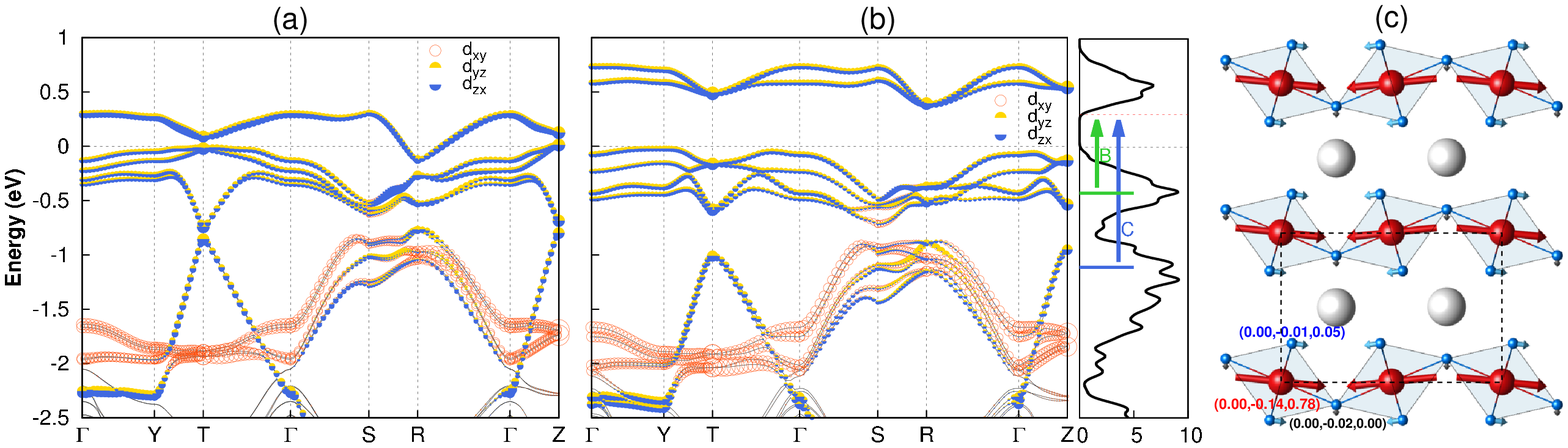} }
\caption{(Color online) Band structure of the AFM structure, obtained using the (a) PBE+SOC and (b) HSE+SOC calculations. The DOS obtained using HSE+SOC is also given in (b). The theoretical estimation of the B and C features observed from RIXS experiment~\cite{Sala} is drawn in DOS. In (c), the sum of the spin and orbital magnetic moments obtained using the HSE+SOC calculation is drawn with three components ($m_a$, $m_b$, $m_c$) along the $a$, $b$, and $c$ axes. Here, $m_i$ is calculated by integrating the corresponding component of magnetic moment inside the PAW sphere with a radius of 1.4 (0.8) {\AA} for Ir (O).}
\end{figure*}

Figure 3(c) shows the sum of the spin and orbital magnetic moments of Ir and O atoms, obtained using HSE+SOC. The total magnetic moment for Ir, O$_1$, and O$_2$ atoms is (0.00, $-$0.14, ${\pm}$0.78), (0.00, $-$0.01, ${\pm}$0.05), and (0.00, $-$0.02, 0.00) in units of ${\mu}_B$, respectively, showing an antiparallel alignment of magnetic moments along the c axis. Here, the Ir magnetic moments are canted along the $b$ axis with ${\sim}$11$^\circ$, comparable with those (2${\sim}$4$^\circ$) reported from RMXS~\cite{Ohgushi} and RIXS~\cite{Sala} experiments.

%\begin{table*}[ht\biguplus]
\begin{table}[ht]
\caption{Spin magnetic moments (in units of ${\mu}_B$) of Ir, O$_1$, and O$_2$ atoms obtained using the PBE, HSE, PBE+SOC, and HSE+SOC calculations. The calculated orbital magnetic moments using PBE+SOC and HSE+SOC are given in parentheses.  O$_1$ is placed in the $ab$ plane while O$_2$ represents the corner-sharing O atom along the $c$ axis (see Fig. 1).}
\begin{ruledtabular}
\begin{tabular}{lccc}
            &  Ir  &  O$_1$  &  O$_2$  \\ \hline
PBE         &  0.31 & 0.05 & 0.00      \\
HSE  & 0.48 & 0.07  &  0.00    \\
PBE+SOC     &  0.16(0.08) & 0.02(0.00) & 0.00(0.00)    \\
HSE+SOC & 0.45(0.34) & 0.05(0.01) & 0.01(0.01)    \\
\end{tabular}
\end{ruledtabular}
\end{table}
%\end{table*}

\begin{figure}[ht]
\centering{ \includegraphics[width=7.7cm]{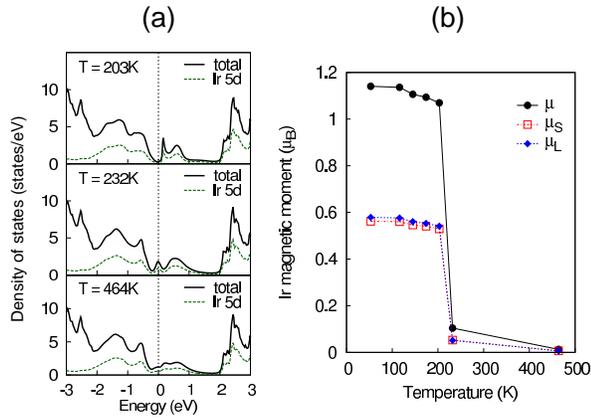} }
\caption{(Color online) (a) Density of states with respect to temperature, obtained using the DFT+DMFT calculation. The calculated magnetic moment (${\mu}$) of Ir atom as a function of temperature
is plotted in (b), together with spin (${\mu}_S$) and orbital (${\mu}_L$) components.}
\end{figure}

Finally, in order to confirm the Slater mechanism for the gap formation in CaIrO$_3$, we perform the DFT+DMFT calculation~\cite{Supple,DMFT} for the AFM and paramagnetic phases. Figure 4(a) shows the DFT+DMFT results for the one-particle spectra. It is seen that the AFM insulating phase is transformed into a paramagnetic metallic phase~\cite{PM} around 230 K. Below this metal-insulator transition (MIT) temperature, the magnetic moment of Ir atom sharply increases [see Fig. 4(b)], indicating that the MIT and the paramagnetic to AFM phase transition are concomitant with each other. The calculated spectral function for the AFM ground state is displayed in Fig. 7S of the Supplemental Material~\cite{Supple}, together with the superimposition of the HSE+SOC band structure. Both results are in good agreement with each other. Therefore, both the comprehensive DFT calculations and the DFT+DMFT calculation reach a consistent conclusion that the gap opening is induced by the AFM order, representing a Slater-type insulator.

In conclusion, our comprehensive DFT calculations with local, semilocal, and hybrid exchange-correlation functionals clarified the effects of tetragonal crystal field, AFM order, and SOC on the $t_{2g}$ states of CaIrO$_3$. We found the large tetragonal crystal field splitting of the $t_{2g}$ states, resulting in the formations of $t_{2g}^{D1}$ doublet (with $d_{yz}$ and $d_{zx}$ characters), $t_{2g}^{D2}$ doublet (with $d_{xy}$ character), and $t_{2g}^{S1}$ and $t_{2g}^{S2}$ singlets (with $d_{yz}$ and $d_{zx}$ characters). We also found that the insulating-gap opening is formed by the interplay between a long-range AFM order and the SOC, representing a spin-orbit Slater insulator. This single-particle based Slater picture is further supported by the DFT+DMFT calculation. Our findings are anticipated to stimulate further experimental and theoretical studies for other iridates such as Sr$_2$IrO$_4$~\cite{BJKim,BJKim2,SJMoon} and Na$_2$IrO$_3$~\cite{Shitade,Comin,Gretarsson} which have been proposed to be a $j_{\rm eff}$ = 1/2 Mott-Hubbard insulator.

\noindent {\bf Acknowledgement.}
This work was supported by National Research Foundation of Korea (NRF) grant funded by the Korea Government (2015R1A2A2A01003248). The calculations were performed by KISTI supercomputing center through the strategic support program (KSC-2014-C3-011) for the supercomputing application research.
Research at Ames Laboratory was supported by the U.S. DOE, Office of Basic Energy Sciences, Division of Materials Sciences and Engineering. Ames Laboratory is operated for the U.S. Department of Energy by Iowa State University under Contract No. DE-AC02-07CH11358.

                  %%%%%  REFERENCES  %%%%%

\noindent Corresponding authors: $^{*}$chojh@hanyang.ac.kr, $^{\dagger}$ykent@iastate.edu

\widetext
\clearpage

\makeatletter
\renewcommand{\fnum@figure}{\figurename ~\thefigure{S}}
\renewcommand{\fnum@table}{\tablename ~\thetable{S}}
\makeatother

\vspace{2.4cm}
{\bf \huge Supplemental Material}
\vspace{0.2cm}

\vspace{1cm}
{\bf \large 1. Calculated structural parameters in the optimized structure of CaIrO$_3$}
\vspace{0.2cm}

\begin{table}[ht]
\caption{Ir-O bond lengths ($d_{1}$ and $d_{2}$ in Fig. 1) and tilt angle (${\theta}_t$ in Fig. 1) obtained using the LDA, PBE, and PBE+SOC calculations, in comparison with the experimental data. For the HSE and HSE+SOC calculations, we used the optimized structures using PBE and PBE+SOC, respectively.}
\begin{ruledtabular}
\begin{tabular}{lccc}
            & $d_{1}$ (\AA)  & $d_{2}$ (\AA) & ${\theta}_t$ ($^\circ$) \\ \hline
LDA         &  1.969 & 2.035 & 22.2    \\
PBE  & 1.973 & 2.049 & 22.5    \\
PBE+SOC  & 1.976 & 2.052 & 22.6    \\
Experiment (Ref.[20])  & 1.972 & 2.049 & 22.3    \\
\end{tabular}
\end{ruledtabular}
\end{table}

\vspace{0.2cm}
{\bf \large 2. Spin-polarized d-orbitals projected band structures}
\vspace{0.2cm}

To understand the microscopic mechanism for the antiferromagnetic (AFM) spin ordering in CaIrO$_3$, we plot the spin-polarized $d$-orbitals projected band structures obtained using the HSE calculation in Fig. 5S. It is seen that the doubly degenerate (spin-up and spin-down) states are localized at Ir$_1$ and Ir$_2$ sites. Since electronic states with the same spin direction can hybridize with each other, the hybridization takes place between the spin-up (or spin-down) states localized at the Ir$_1$ and Ir$_2$ sites, yielding not only an energy gain from the exchange kinetic energy but also the formation of hybridization gap (E$_{g}^{S1-S2}$) or the enhanced energy splitting. This so-called superexchange interaction between Ir$^{4+}$ spins results in the stabilization of the AFM order along the $c$ axis.

\begin{figure}[h]
\centering{ \includegraphics[width=9.5cm]{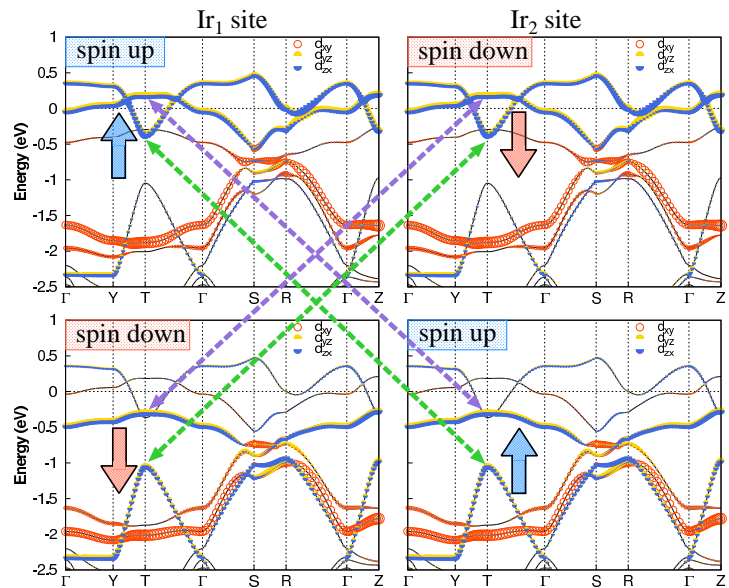} }
\caption{Spin-polarized d-orbitals projected band structures of the AFM structure obtained using the HSE calculation. The bands projected onto $d_{xy}$, $d_{yz}$, and $d_{zx}$ orbitals are displayed with circle and semicircles whose the radii are proportional to the weights of corresponding orbitals. Ir$_1$ and Ir$_2$ are indicated in the Fig. 1 of the main text. The energy zero represents the Fermi level.
}
\end{figure}

\newpage
{\bf \large 3. Partial densities of states of the $d_{xy}$, $d_{yz}$, and $d_{zx}$ orbitals}
\vspace{0.2cm}

We plot the partial densities of states (PDOS) of the $d_{xy}$, $d_{yz}$, and $d_{zx}$ orbitals using the HSE and HSE+SOC calculation in Fig. 6S. For the $t_{2g}$ state above $E_F$, the PDOS obtained using HSE give a ratio of 0\%:50\%:50\%, but they change into 8\%:46\%:46\% for the HSE+SOC calculation. The latter PDOS ratio of the $d_{xy}$, $d_{yz}$, and $d_{zx}$ orbitals is very consistent with the interpretation of RIXS experiment [13] that the $t_{2g}$ state near $E_F$ was estimated as ${\mp}0.32|xy,{\mp}>$ + 0.67 (($|yz,{\pm}>{\mp}i|zx,{\pm}>$).

\begin{figure}[h]
\centering{ \includegraphics[width=14.0cm]{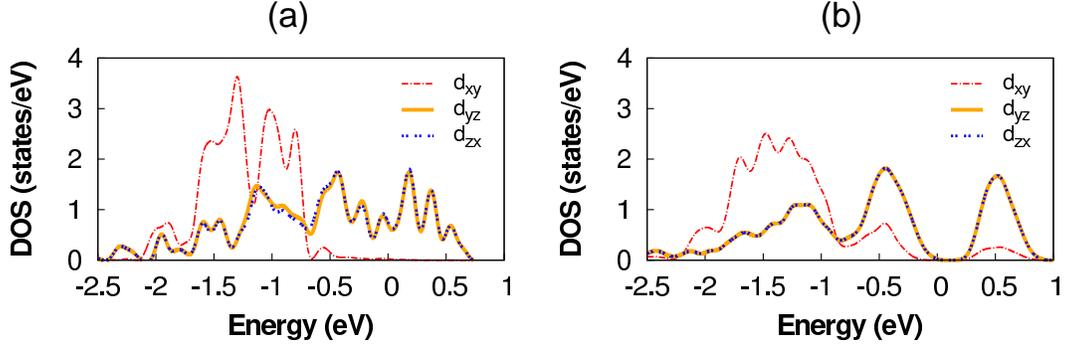} }
\caption{PDOS of the $d_{xy}$, $d_{yz}$, and $d_{zx}$ orbitals obtained by the (a) HSE and (b) HSE+SOC calculation. }
\end{figure}

\vspace{2cm}

{\bf \large 4. Computational details of the DFT+DMFT calculation}
\vspace{0.2cm}

The dynamical mean field theory (DMFT) calculations are performed in a fully charge self-consistent way
by combining with density functional theory (DFT) using the projection-embedding implementation [22, 35].
The DFT calculations within the generalized gradient approximation (GGA) [23] are carried out
using the full-potential linearized augmented plane-wave (FLAPW) method
as implemented in the WIEN2K package [24].
In the DMFT part, we solve the quantum impurity model using the continuous-time quantum Monte Carlo method [25, 26].
The local correlated orbitals are projected from a 20eV energy window around the Fermi level.
The double-counting functional is expressed as $\Phi_{dc}(n_d)=n_dE_{dc}$,
where $E_{dc}=U(n_d^0-1/2)-J/2(n_d^0-1)$ and the nominal occupancy $n_d^0$ of Ir ion in CaIrO$_3$ is $5$.
Due to the large energy window, the local Coulomb interaction parameters are relatively system-independent
and estimated [27] as $U\approx 4.5$ eV and $J\approx 0.8$ eV [28].
We set different local coordinates on each Ir atom in order to simulate the antiferromagnetic structure.
With the crystal-field and spin-orbit coupling interactions considered,
the effective $J_{\text{eff}}=1/2$ states are constructed as,
\begin{equation}
|\psi_{\pm\frac{1}{2}}\rangle =\mp\sqrt{\frac{3-2\gamma(\omega)^2}{3}}|d_{xy},\mp\sigma\rangle
+ \frac{\gamma(\omega)}{\sqrt{3}}(|d_{yz},\pm\sigma\rangle \mp i |d_{xz},\pm\sigma\rangle)
\end{equation}
where $\sigma$ labels the spin [28].
The value of $\gamma$ at low-frequency limit is adopted in our transformation in which $\gamma(\omega=0)\approx 1.16$.
Then the magnitudes of magnetic moments along the magnetization axis can be calculated as,
$\langle \mu_L^{z} \rangle=2\gamma^2\Delta n/3$ and $\langle \mu_S^{z}\rangle = (4\gamma^2-3)\Delta n/3$,
where $\Delta n$ is the difference of the occupation numbers of $|\psi_{\pm1/2}\rangle$ [28].

\vspace{2cm}

\newpage

{\bf \large 5. Spectral function of the AFM state obtained using DFT+DMFT}
\vspace{0.2cm}

The calculated spectral function of the AFM ground state is displayed in Fig. 7S, together with the superimposition of the HSE+SOC band structure. Both results are in good agreement with each other.

\begin{figure}[h]
\centering{ \includegraphics[width=11.0cm]{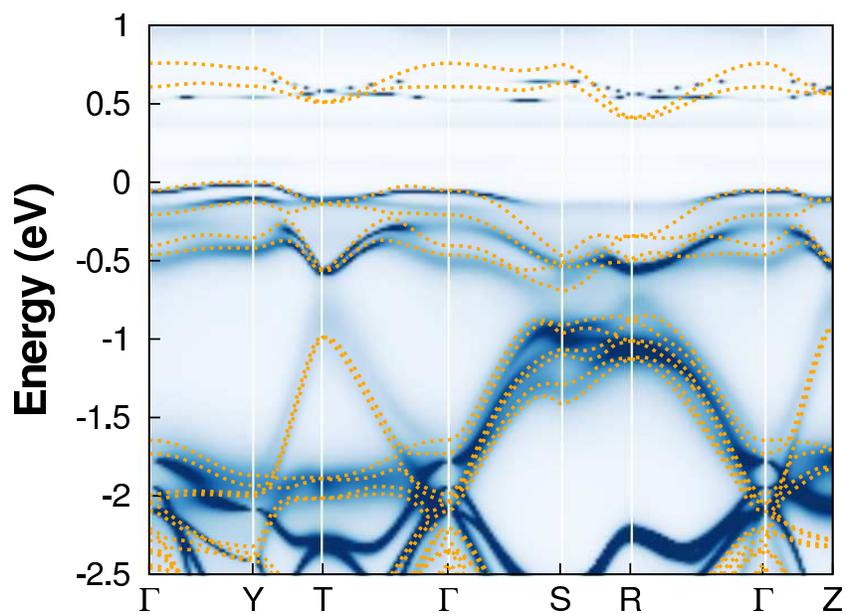} }
\caption{Spectral function of the AFM state obtained using DFT+DMFT. Dotted lines denote the band structure obtained using HSE+SOC. Here, the DFT+DMFT spectral function and the HSE+SOC band structure are aligned to have the same position at the topmost valence state at $Y$.}
\end{figure}


\begin{thebibliography}{99}
\bibitem{DOS}
For intra-$t_{2g}$ excitations, the measured RIXS spectra were fitted to three Pearson functions in the energy range between 0.3 and 1.6 eV below the conduction band edge: i.e., a weak, energy-resolution limited peak (A) with an excitation energy of ${\sim}$0.42 eV and two intense broad features (B and C) centered at 0.65 and 1.22 eV. In order to compare these RIXS excitations with our calculated $t_{2g}$ bands, we plot the density of states (DOS) obtained using HSE+SOC in Fig. 3(b). We find that the calculated DOS exhibits two well-separated energy regions below $E_F$: i.e., the region I from $E_F$ to $-$0.8 eV and the region II from $-$0.8 to $-$1.3 eV. Here, since the intra-$t_{2g}$ excitations in RIXS spectra were measured in the energy range between 0.3 and 1.6 eV below the conduction band edge, we take the energy range of the region II from $E_{F}-$0.8 to $E_{F}-$1.3 eV. We estimate the center of each region as ${\epsilon}_{\rm I}$ = $-$0.45 eV and ${\epsilon}_{\rm II}$ = $-$1.13 eV by using a relation of ${\epsilon}$ = ${\int}$ DOS($E$)${\times}E$ d$E$ / ${\int}$ DOS($E$) d$E$. Considering that the HSE+SOC band gap is 0.32 eV, the excitation energies of two $d$-$d$ transitions are estimated to be 0.77 and 1.45 eV, well comparable with those (0.65 and 1.22 eV) for the B and C features in RIXS experiment. We note that, although the observed A peak cannot be clearly identified from the calculated DOS, it may be associated with the flat $t_{2g}$ bands near $E_F$ [see Fig. 3(b)].
\end{thebibliography}
\end{document}